\documentclass[aps,twocolumn,prb,groupedaddress,amsmath,amssymb,10pt,floatfix]{revtex4-2}
\usepackage{dcolumn}
\usepackage{bm}
\usepackage{amsmath}
\usepackage{txfonts}
\usepackage[T1]{fontenc}
\usepackage{xspace}
\usepackage{ulem}
\usepackage{comment}
\usepackage{braket}
\usepackage{bm}
\usepackage{physics}

\ifx\pdfoutput\undefined
\usepackage{graphicx}
\usepackage{hyperref}
\usepackage{color}
\else
\usepackage{graphicx}
\usepackage{hyperref}
\usepackage{color}
\fi

\begin{document}
\let\emph\textit
\title{Generalized Nagaoka ferromagnetism accompanied by flavor-selective Mott states in an SU($N$) Fermi-Hubbard model}

\author{Juntaro Fujii}
\email{fujii.j.47dc@m.isct.ac.jp}
\author{Kazuki Yamamoto}
\author{Akihisa Koga}
\affiliation{
Department of Physics, Institute of Science Tokyo, Meguro, Tokyo 152-8551, Japan
}

\date{\today}

\begin{abstract}
  We study the ferromagnetic instability
  in an $\mathrm{SU}(N)$ Fermi-Hubbard model on the hypercubic lattice.
  Combining dynamical mean-field theory
  with continuous-time quantum Monte Carlo simulations,
  we find that, in the strong-coupling regime at low temperatures,
  ferromagnetically ordered (FM) states develop
  away from the commensurate fillings.
  In the particle-doped SU($3$) system near one-third filling,
  the FM state is accompanied by a spontaneous flavor-selective Mott state,
  where two of the three flavors are Mott insulating while the remaining flavor is metallic.
  Since particles in the metallic flavor can almost freely move on the lattice
  without correlation effects,
  the ordered state is stabilized by the kinetic-energy gain of the doped particles.
  This is similar to the generalized Nagaoka ferromagnetism
  discussed in the one-hole-doped system at one-third filling.
  In the SU($4$) case, we find that six distinct types of FM states
  appear as the particle density varies.
  The results uncover the nature of 
  the FM state in the SU($N$) Fermi-Hubbard systems and 
  highlight the rich magnetic behavior enabled by enlarged internal symmetries.
\end{abstract}

\maketitle

\section{Introduction}\label{sec:i}
Recent experimental progress with ultracold atoms has enabled the exploration of strongly correlated quantum phenomena~\cite{takahashi_2020,greif2013short,hart2015observation,boll2016spin,parsons2016site,cheuk2016observation,he2019recent,lebrat2024observation}, by utilizing Feshbach resonances~\cite{Chin10} and optical lattices~\cite{grimm2000optical}.
Fermionic systems with multiple internal states 
have been realized by exploiting the hyperfine states of atoms 
such as $^{40}$K~\cite{Demarco1999}, 
$^6$Li~\cite{Truscott2001,Schreck2001,Granade2002}, 
$^{173}$Yb~\cite{Fukuhara2007degenerate,taie2010realization,pasqualetti2024EOS,pagano2014one},
$^{171}$Yb~\cite{taie2010realization}, and
$^{87}$Sr~\cite{desalvo_degenerate_2010,zhang2014spectroscopic}.
While SU($2$) systems have been extensively investigated,
these multicomponent atomic gases now enable the study of SU($N$) Fermi-Hubbard models
with $N>2$~\cite{hubbard1963electron,carsten2004ultracold},
where the metal-insulator crossover~\cite{hofrichter_direct_2016} and antiferromagnetic correlations~\cite{ozawa_antiferromagnetic_2018,taie_observation_2022} have been observed.
Generalizing strongly correlated phenomena to those with $\mathrm{SU}(N)$ symmetry 
has opened up new possibilities for understanding exotic quantum phases in condensed matter physics~\cite{congjun2003exact,carsten2004ultracold,Michael2009Mott,Rapp2011ground_state,cazalilla2014ultracold,capponi_phases_2016,Yoshida2021rigorous,yamamoto_universal_2023,Feng_2023,Nakagawa2024exact,ibarra_many_body_su_N_2024}.

Ferromagnetism in correlated electron systems has long been a central topic in condensed matter physics.
The ferromagnetically ordered (FM) states are often realized in realistic multiorbital systems,
originating from ferromagnetic exchange mechanisms such as Hund's coupling and double exchange.
By contrast, it remains unclear
whether the FM state can emerge
in the Fermi-Hubbard model without explicit magnetic exchange interactions.
In the weak-coupling regime, the Stoner criterion~\cite{stoner1938collective,hubbard1963electron} and Slater-type ferromagnetism~\cite{slater1936ferromagnetism} have been discussed,
although their applicability is limited.
Flat-band ferromagnetism has also been studied as another route toward itinerant magnetism~\cite{mielke1991ferromagnetism,ferromagnetism1992tasaki}. 
However, few theoretical approaches are available in the strong-coupling regime,
and systematic understanding remains limited.
A prominent exception is Nagaoka ferromagnetism,
which is rigorously established for the single-hole-doped SU($2$) Fermi-Hubbard model at half filling~\cite{nagaoka_1966}.
Its extensions to finite hole doping, finite temperatures, and higher symmetries
such as SU($N$) have been discussed by numerical calculations~\cite{obermeier_ferromagnetism_1997,zitzler_magnetism_2002,park_dynamical_2008,kamogawa_ferromagnetic_2019,fujii2025itinerant,samajdar_polaronic_2024,huang_2023,sharma2025nagaoka,jin2026flat} as well as analytical studies~\cite{cazalilla_2009,katsura_2013,singh_divergence_2022,kim2024itinerant}.
In our previous work~\cite{fujii2025itinerant},
by means of reliable numerical methods,
we clarified that an FM state characterized by the generalized Nagaoka ferromagnetism emerges in the SU($3$) Fermi-Hubbard model when holes are doped away from one-third filling. 
In general, for $N>2$, the internal flavor degrees of freedom introduce many competing configurations
even at commensurate fillings~\cite{Sotnikov2014magntic}, and the kinetic-energy-driven mechanism
underlying Nagaoka ferromagnetism does not directly carry over.
It still remains unclear whether FM phases can emerge
around commensurate fillings in SU($N$) systems beyond the Nagaoka mechanism.
Therefore, detailed numerical investigations are essential for clarifying
the ferromagnetic instability in the SU($N$) Fermi-Hubbard model.

In this paper, we investigate the ferromagnetic instability
in the $\mathrm{SU}(N)$ Fermi-Hubbard model on the hypercubic lattice,
using dynamical mean field theory (DMFT)~\cite{metzner_correlated_1989,muller_hartmann_correlated_1989,georges_dynamical_1996}.
In the $\mathrm{SU}(3)$ case,
we demonstrate that,
in addition to the well-known FM state, which is adiabatically connected to the generalized Nagaoka ferromagnetism,
another FM phase emerges together with
a spontaneous flavor-selective Mott state above one-third filling,
where two of the three flavors are Mott insulating and
the other is metallic. 
For the $\mathrm{SU}(4)$ case, 
six distinct types of FM states are realized
away from the commensurate fillings.
These results highlight how increasing the number of internal components qualitatively
enriches the magnetic phase structure.
The ferromagnetic instability in the SU($N$) Fermi-Hubbard model
is also addressed.

The rest of this paper is organized as follows. 
In Sec.~\ref{sec:mm}, we introduce the $\mathrm{SU}(N)$ Fermi-Hubbard model
and briefly describe the DMFT framework.
In Sec.~\ref{sec:r}, we clarify that,
in the strong-coupling regime at low temperatures,
the FM states
are stabilized in the SU($3$) and SU($4$) Fermi-Hubbard models.
The ferromagnetic instability in the SU($N$) case is also discussed.
Conclusions are given in Sec.~\ref{sec:conclusion}.

\section{Methods}\label{sec:mm}
We consider the $\mathrm{SU}(N)$ Fermi-Hubbard model 
on a hypercubic lattice, 
which is described by the Hamiltonian
\begin{equation}
  \hat{H} = 
  -t \sum_{\langle i,j \rangle} \sum_{\sigma=1}^{N} 
  \hat{c}_{i,\sigma}^{\dagger}\hat{c}_{j,\sigma}
  + \frac{U}{2} \sum_{i} \sum_{\sigma \neq \sigma'} 
  \hat{n}_{i,\sigma} \hat{n}_{i,\sigma'},
\label{hamiltonian}
\end{equation}
where $\hat{c}_{i,\sigma}^{\dagger}$ ($\hat{c}_{i,\sigma}$) 
is the creation (annihilation) operator 
of a fermion with flavor $\sigma~(=1, 2, \ldots, N)$ at site $i$, 
and $\hat{n}_{i,\sigma} = \hat{c}_{i,\sigma}^{\dagger}\hat{c}_{i,\sigma}$. 
$t$ is the hopping amplitude between nearest-neighbor sites
$\langle i,j\rangle$ and $U$ is the on-site repulsive interaction.
We employ the grand canonical ensemble, in which  
the expectation value of an operator $\hat{A}$ is given by
$\expval{\hat{A}} = \Tr \left[e^{-(\hat{H}-\mu \hat{N})/T} \hat{A}\right]/Z$,
where $\mu$ is the chemical potential, $T$ is the temperature,
$Z=\Tr \left[e^{-(\hat{H}-\mu \hat{N})/T}\right]$, and
$\hat{N} = \sum_{i,\sigma} \hat{n}_{i,\sigma}$.
The particle density of each flavor is defined 
by $n_{\sigma} = \sum_{i} \expval{\hat{n}_{i,\sigma}}/L$, 
where $L$ is the number of lattice sites.
The total number of particles is given by
$n_{\mathrm{tot}} = \sum_{\sigma} n_{\sigma}$.
Since the model is particle-hole symmetric,
we focus on the filling range $0\le n_{\mathrm{tot}} \le N/2$.

To study low-temperature properties in the SU($N$) Fermi-Hubbard model, we employ DMFT~\cite{metzner_correlated_1989,muller_hartmann_correlated_1989,georges_dynamical_1996}.
In DMFT, the original lattice model 
is mapped to a single-impurity model coupled to an effective bath. 
It is known that, in the limit of infinite spatial dimensions,
DMFT is exact~\cite{georges_dynamical_1996}, as
it correctly captures local dynamical correlations.
DMFT has been successfully applied 
to analyze various strongly correlated phenomena such as 
the Mott transition~\cite{Kotliar_1996,Rozenberg_1997,Han_1998,Koga_2002,Ono_2003,koga2004orbital,koga_2005}, 
superconductivity~\cite{Capone_2001,Rapp_2007,Inaba_2009,Hoshino2015supercunductivity,koga_2015,ishigaki_2018,Okanami2014Stability,Yue_2021}, 
and magnetic ordering~\cite{Momoi_1998,Held_1998,yanatori_finite_2016,Koga_2017}.
In general, magnetically ordered states with two or more sites in the unit cell
may be realized in the SU($N$) Fermi-Hubbard model.
In this paper, we focus on the ferromagnetic instability
in the strong-coupling regime
by restricting our discussion to spatially homogeneous magnetic order.

The FM state we focus on in this study is generally
characterized by the order parameter $m_\alpha=\langle \hat{S}_\alpha\rangle/L$ with
\begin{align}
  \hat{S}_\alpha &=
  \frac{1}{2} \sum_{i,\sigma,\sigma'} 
  \hat{c}_{i,\sigma}^{\dagger} 
  (\lambda_{\alpha})_{\sigma,\sigma'} 
  \hat{c}_{i,\sigma'}, 
\end{align}
where $\lambda_{\alpha}$ $(\alpha=1,\ldots, N^2-1)$
are the generators of
SU($N$) with ${\rm Tr}[\lambda_\alpha\lambda_\beta]=2 \delta_{\alpha,\beta}$~\cite{georgi2000lie,gell_mann_symmetries_1962}.
Since imbalance in the flavor occupations serves as
a reliable indicator of ferromagnetic order,
it is sufficient to characterize the magnetic order
using the $(N-1)$ diagonal generators, whose matrix elements have vanishing off-diagonal components.
Consequently, one can restrict the analysis to the diagonal components of Green's function
without loss of generality.

The Dyson equation for the lattice Green's function 
is then given by
\begin{equation}
  G^{\sigma}(k,i\omega_n)^{-1}
  = i\omega_n + \mu - \epsilon_k - \Sigma^{\sigma}(k,i\omega_n),
\end{equation}
where $\epsilon_k$ denotes the dispersion relation,
$\omega_n [= (2n+1)\pi T]$ is the Matsubara frequency, and 
$G^{\sigma}(k,i\omega_n)$ and $\Sigma^{\sigma}(k,i\omega_n)$ 
represent the lattice Green's function and the self-energy with flavor $\sigma$.
In infinite dimensions, the self-energy is momentum-independent,
$\Sigma_{\mathrm{loc}}^{\sigma}(i\omega_n) = \Sigma^{\sigma}(k,i\omega_n)$,
and the local Green's function is given by
\begin{equation}
  G_{\mathrm{loc}}^{\sigma}(i\omega_n)
  = \int \frac{\rho(\epsilon)\, d\epsilon}
  {i\omega_n + \mu - \epsilon - \Sigma_{\mathrm{loc}}^{\sigma}(i\omega_n)},
\end{equation}
where $\rho(\epsilon)$ denotes the noninteracting density of states (DOS). 
For the hypercubic lattice, 
the noninteracting DOS is given by
\begin{equation}
  \rho(\epsilon)
  = \frac{1}{\sqrt{\pi} D}
    \exp\!\left[-\left(\frac{\epsilon}{D}\right)^{2}\right],
\end{equation}
where $D$ is the characteristic energy.
In the effective impurity model, 
the Weiss function is given by
\begin{equation}
  \mathcal{G}^{\sigma}(i\omega_n)^{-1} 
  = i\omega_n + \mu - \Delta^{\sigma}(i\omega_n),
  \label{weiss_function}
\end{equation}
where $\Delta^{\sigma}(i\omega_n)$ denotes the hybridization function~\cite{georges_dynamical_1996}.
The self-consistent equations in the framework of DMFT are given by
$G_{\mathrm{imp}}^{\sigma}(i\omega_n) = G_{\mathrm{loc}}^{\sigma}(i\omega_n)$ 
and $\Sigma_{\mathrm{imp}}^{\sigma}(i\omega_n) = \Sigma_{\mathrm{loc}}^{\sigma}(i\omega_n)$.
For a given Weiss function, we solve the effective impurity model to compute
the impurity Green's function $G_{\mathrm{imp}}^{\sigma}(i\omega_n)$ and
self-energy $\Sigma_{\mathrm{imp}}^{\sigma}(i\omega_n)$,
obtain the lattice Green's function $G_{\mathrm{loc}}^{\sigma}(i\omega_n)$,
update the Weiss function,
and iterate this self-consistent procedure
until convergence is reached within numerical accuracy.

In this study,
we employ the hybridization-expansion CTQMC method~\cite{werner_continuous_2006}
as the impurity solver to systematically investigate low-temperature properties.
Since we focus on the region $U\gg D$ and $T\ll D$,
we adopt two improvements to enhance efficiency and accuracy.
First, the double-flip update~\cite{koga2011polarized} is employed
to maintain a reasonable acceptance ratio in the strong-coupling regime.
Second, we use the nonuniform sampling scheme~\cite{fujii2025itinerant} together with the intermediate representation basis~\cite{shinaoka_compressing_2017,chikano_irbasis_2019},
which allows for an accurate representation of the Green's function.
These improvements are essential for quantitative analysis of
the ferromagnetic instability at low temperatures
in the strong-coupling regime.

In this calculation,
we evaluate several physical quantities.
The magnetic susceptibility is defined as 
\begin{equation}
  \chi_\alpha = \lim_{h\rightarrow 0} \frac{m_\alpha}{h_\alpha},
\end{equation}
where $h_\alpha$ is an external field that couples to the system via the Zeeman Hamiltonian $-h_\alpha \hat{S}_\alpha$.
We note that
the magnetic susceptibility is independent of $\alpha$ (see Appendix~\ref{ap:iso} for details).
For simplicity, we evaluate it using one of the diagonal generators of SU($N$), $\{\lambda_\alpha\}$, whose diagonal part is represented by the flavor-imbalance vector
\begin{equation}
\bm{v}_N=\sqrt{\frac{2}{N(N-1)}}\Big(1,1,\cdots, -(N-1)\Big). 
\label{eq_vn}
\end{equation}
In this case, the magnetization is given by $m=\sqrt{(N-1)/(2N)}(n_1-n_N)$.
To discuss the stability of the FM state,
we also evaluate the kinetic energy per site 
defined by $K=\sum_\sigma K_\sigma$ with
$K_{\sigma} = -\int_0^{\beta} d\tau\, G_{\mathrm{imp}}^{\sigma}(\tau) \Delta^{\sigma}(-\tau)$
and the interaction energy $I=U\sum d_{\sigma,\sigma'}/2$,
where $d_{\sigma,\sigma'}[=\sum_i \langle \hat{n}_{i,\sigma} \hat{n}_{i,\sigma'} \rangle/L]$
is the double occupancy between flavors $\sigma$ and $\sigma'$.
In addition, we consider the following quantity
\begin{align}
  A_\sigma= -\frac{1}{\pi T} G_{\mathrm{imp}}^{\sigma}\left(\frac{1}{2T}\right),
\end{align}
which, at zero temperature, reduces to the DOS at the Fermi
level~\cite{trivedi1996superconductor,tong2001mott,vanLoon2014beyond,goldberger2024dynamical}.
Thus, even at finite temperatures,
$A_\sigma$ allows us to discuss whether each flavor exhibits metallic or insulating behavior.
In the following, we discuss the ferromagnetic instability
in the SU($N$) Fermi-Hubbard model by examining several physical quantities.
\section{Numerical Results}\label{sec:r}

\subsection{Ferromagnetism in the SU($3$) Fermi-Hubbard model}\label{ss:su3}
\begin{figure}
\begin{center}
\includegraphics[width=\linewidth]{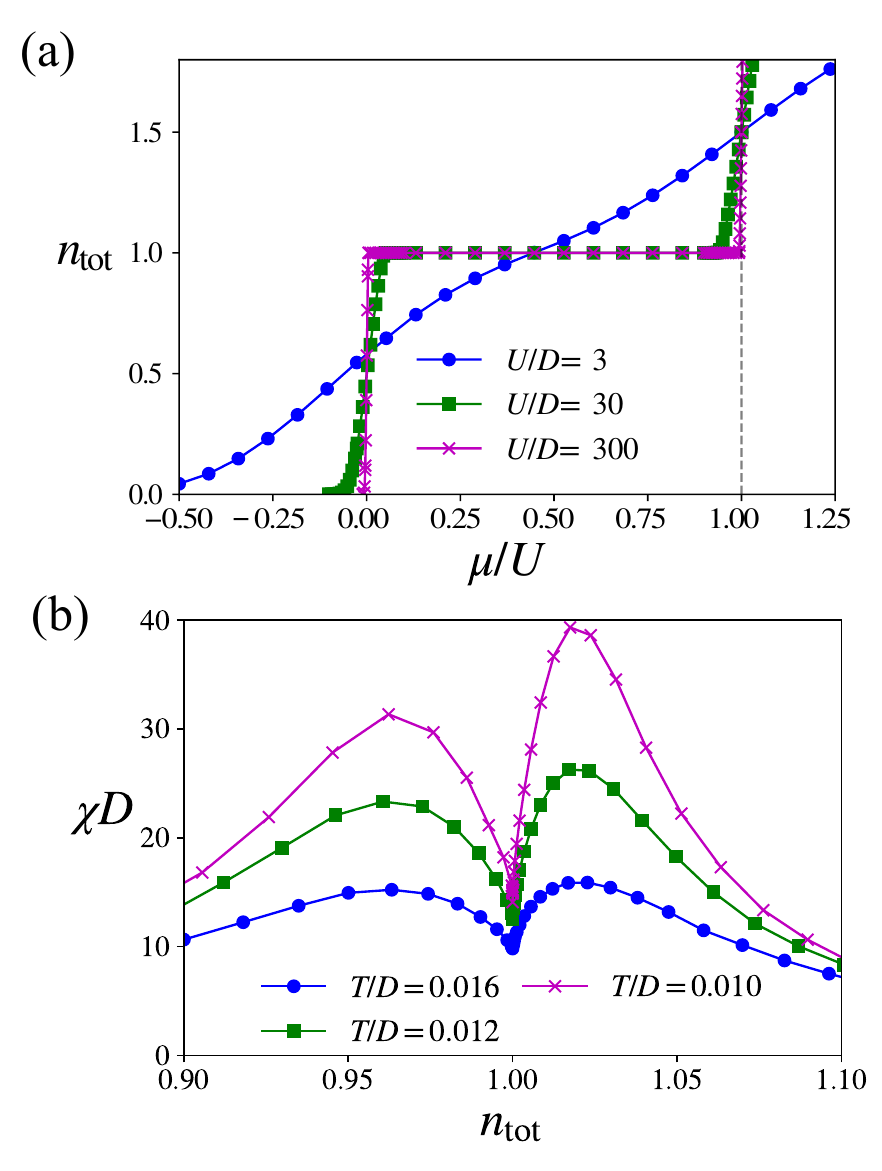}
\end{center}
\caption{
(a) Total number of particles $n_{\mathrm{tot}}$ 
as a function of the chemical potential
for the $\mathrm{SU}(3)$ Fermi-Hubbard model 
at $T/D = 0.01$
when $U/D = 3$ (blue), $30$ (green), and $300$ (magenta).
The dashed line indicates the chemical potential corresponding to the half-filled condition.
(b) Magnetic susceptibility
as a function of $n_{\mathrm{tot}}$
for $U/D = 300$ when $T/D=0.010$ (blue), $0.012$ (green), and $0.016$ (magenta).
}
\label{fig_A_v1}
\end{figure}

We consider the SU($3$) Fermi-Hubbard model
to discuss its ferromagnetic instability.
First, we examine the chemical potential dependence
of the total number of particles.
The results for $T/D = 0.01$ are shown in Fig.~\ref{fig_A_v1}(a).
We find that, as the chemical potential increases, 
$n_{\mathrm{tot}}$ smoothly increases in the weak-coupling case ($U/D = 3$), 
while a plateau appears at $n_{\mathrm{tot}} = 1$
in the strong-coupling cases ($U/D=30$ and $300$).
This plateau signals the formation of 
a Mott insulating phase at one-third filling~\cite{cazalilla2014ultracold}.
In contrast, for the intermediate filling regime with $n_{\mathrm{tot}} \neq 1$, 
the filling varies smoothly with increasing $\mu$, 
implying that metallic states are realized away from the commensurate filling. 

To study the ferromagnetic instability in the system, 
we examine the magnetic susceptibility
since its temperature dependence is useful for
identifying precursors of ferromagnetism~\cite{kamogawa_ferromagnetic_2019,fujii2025itinerant}.
In Fig.~\ref{fig_A_v1}(b), we show the magnetic susceptibility
for $U/D = 300$ when $T/D=0.010, 0.012$, and $0.016$.
We find two peaks
below and above the one-third filling, $n_{\mathrm{tot}}=1$.
A notable feature is that the magnetic susceptibility
around these peaks is significantly enhanced as the temperature decreases.
The lower-density peak ($n_{\mathrm{tot}}\sim 0.96$) reflects
the well-established precursor of generalized Nagaoka ferromagnetism,
which has been investigated in detail in Ref.~\cite{fujii2025itinerant}.
In addition, similar behavior appears in the higher-density regime ($n_{\mathrm{tot}} \sim 1.02$),
suggesting the formation of the FM state at low temperatures.
In contrast, in the dilute limit ($n_{\mathrm{tot}} \sim 0$), 
we find no peak singularity (not shown).
This means that the ground state should be paramagnetic
since particle correlations have little effect on the system.

\begin{figure}
\begin{center}
\includegraphics[width=\linewidth]{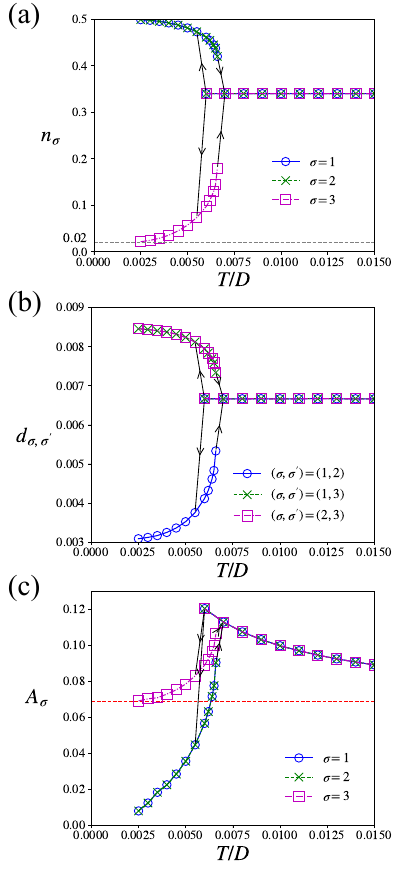}
\end{center}
\caption{
(a) Particle density $n_\sigma$, (b) double occupancy $d_{\sigma,\sigma'}$, 
and (c) the quantity $A_\sigma$ in the $\mathrm{SU}(3)$ Fermi--Hubbard model
with $U/D = 300$ and $n_{\mathrm{tot}} = 1.02$.
The red dashed line in (c) represents the value of the noninteracting 
DOS corresponding to $n_3 = 0.02$.
Black arrows indicate the evolution of the DMFT solutions
upon varying the temperature $T$,
highlighting the coexistence of FM and PM solutions.
}
\label{fig_CDE}
\end{figure}
To clarify whether the FM state is indeed realized around $n_{\mathrm{tot}}\sim 1.02$,
we calculate the temperature dependence of
the particle density for each flavor for $U/D=300$,
as shown in Fig.~\ref{fig_CDE}(a).
At higher temperatures,
the paramagnetic (PM) state is realized
with equal particle densities $n_1=n_2=n_3$.
In contrast, we find a clear density imbalance at low temperatures, 
which indicates spontaneous symmetry breaking.
Since $n_1 = n_2 \neq n_3$, the FM state is characterized by the flavor-imbalance vector $\bm{v}_3$ defined in Eq.~\eqref{eq_vn}.
As the temperature increases,
the imbalance in particle density decreases and suddenly vanishes at $T=T_{c_2}(\sim 0.0066D)$.
This implies that the first-order magnetic phase transition to the PM state occurs,
which is essentially the same as the ferromagnetic transition in the hole-doped case~\cite{fujii2025itinerant}.
The PM solution survives down to $T=T_{c_1}(\sim 0.0055D)$.

One of the most important points is that $n_1=n_2$ approaches $0.5$
with decreasing temperature,
which suggests that a Mott insulating state is realized
for flavors $\sigma=1$ and $2$. 
To confirm this,
we examine the double occupancy $d_{\sigma,\sigma'}$ and the quantity $A_{\sigma}$,
as shown in Figs.~\ref{fig_CDE}(b) and \ref{fig_CDE}(c).
The latter may be regarded as the DOS at the Fermi level.
As the temperature decreases,
we find $d_{12} \ll d_{23}=d_{31}$.
Furthermore, we find that $A_{1}$ and $A_{2}$ approach zero, 
whereas the other remains finite. 
These results indicate that 
flavors $\sigma = 1, 2$ become Mott insulating, 
while flavor $\sigma = 3$ remains metallic. 
This demonstrates that the ferromagnetic order is accompanied 
by a spontaneous flavor-selective Mott state~\cite{koga2004orbital,deMedici2005orbital,Rapp2011ground_state,hoshino2017spontaneous}. 
This naturally suggests that, in the zero-temperature limit,
one particle occupies either flavor 1 or 2 at each site, while
the remaining particles move freely in the other flavor.
To further clarify this, the noninteracting DOS at the Fermi level
for the particle density $n_3=0.02$ is shown as a red dashed line in Fig.~\ref{fig_CDE}(b).
We find that $A_3$ asymptotically approaches this value with decreasing temperature,
providing additional evidence that, in the FM phase,
particles with flavor $\sigma=3$ are effectively noninteracting
whereas particles with the other flavors are completely localized.
This conclusion is further supported by analyzing the energy gain,
whose dominant contribution arises from the kinetic energy of flavor $\sigma=3$ (not shown).

\begin{figure}
\begin{center}
\includegraphics[width=\linewidth]{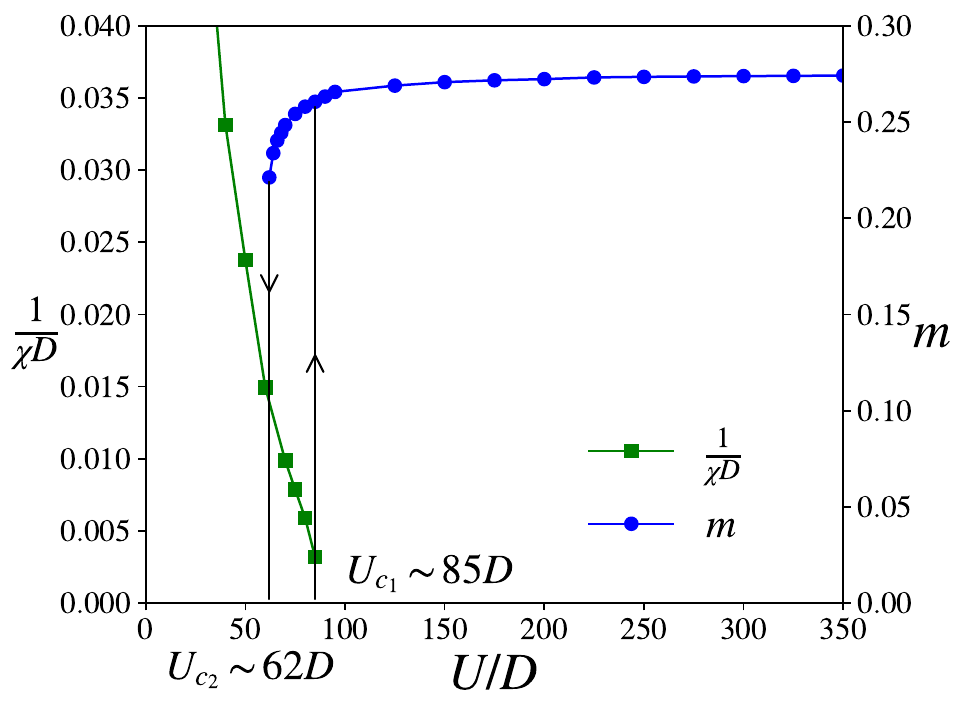}
\end{center}
\caption{
Magnetization $m$ (right axis) 
and inverse magnetic susceptibility $1/\chi D$ (left axis) 
as functions of the interaction strength $U/D$ 
for the $\mathrm{SU}(3)$ Fermi-Hubbard model
at $T/D = 0.003$ and $n_{\mathrm{tot}} = 1.02$.
Black arrows indicate the evolution of the DMFT solutions
upon varying the interaction strength $U$,
highlighting the coexistence of FM and PM solutions.
}
\label{fig_P_v1}
\end{figure}

We also discuss the stability of the FM state
with respect to interaction strength.
The magnetization $m$ and the inverse susceptibility $1/(\chi D)$
for $T/D = 0.003$ and $n_{\mathrm{tot}} = 1.02$ are shown in Fig.~\ref{fig_P_v1}.
When $U$ is small, the PM state is realized
with $n_\sigma=n_{\mathrm{tot}}/3$ and $m=0$.
The susceptibility increases with increasing interaction strength.
The PM solution suddenly vanishes at $U=U_{c_1} \;(U_{c_1}/D \sim 85)$, and
spontaneous magnetization emerges.
A further increase in the interaction strength leads to a
monotonic increase in the magnetization.
This behavior contrasts with that of the antiferromagnetically ordered state
stabilized by
the Heisenberg interaction $J\sim D^2/U$,
where the magnetization is suppressed in the strong-coupling regime at finite temperatures.
When the interaction strength decreases from the strong-coupling regime, 
the magnetization abruptly vanishes at $U=U_{c_2} \;(U_{c_2}/D \sim 62)$. 
Therefore, the system undergoes a first-order magnetic phase transition accompanied by hysteresis. 
In the region $U_{c_2}< U < U_{c_1}$, 
both FM and PM solutions coexist.
\begin{figure}
\begin{center}
\includegraphics[width=\linewidth]{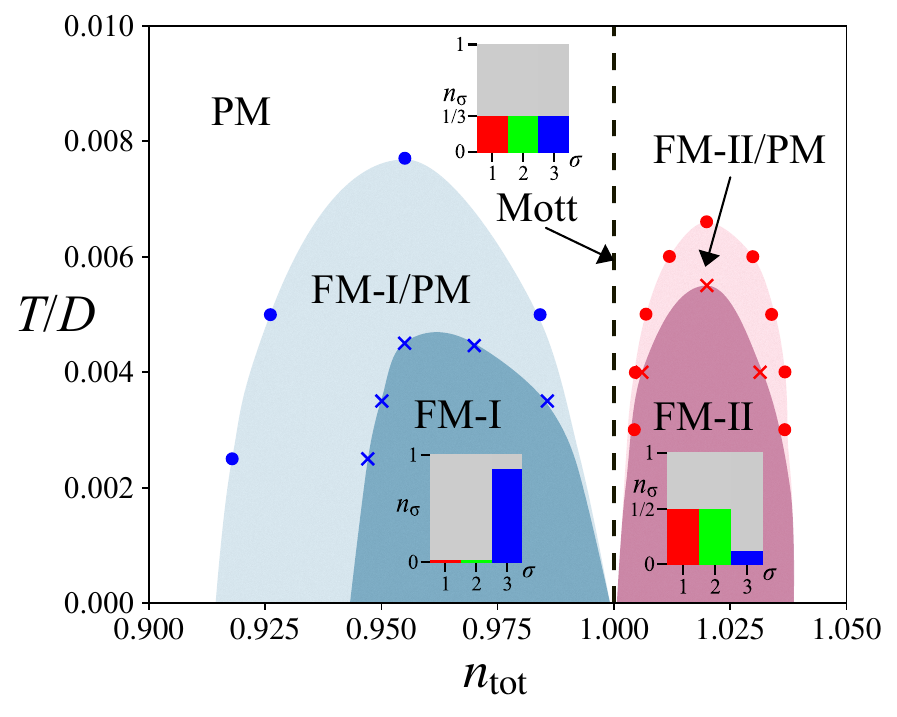}
\end{center}
\caption{
  Finite-temperature phase diagram of the $\mathrm{SU}(3)$ Fermi-Hubbard model
  for $U/D = 300$ around one-third filling $(n_{\mathrm{tot}}\sim 1)$.
  Blue (red) circles and crosses represent the phase transition points, where
  FM-I (FM-II) and PM states disappear, respectively.
  The Mott insulating state at one-third filling $n_{\mathrm{tot}}=1$ is 
  indicated by the dashed line. 
  The flavor occupancies for FM-I, FM-II, and Mott states are illustrated.
}
\label{fig_G_v2}
\end{figure}

We show in Fig.~\ref{fig_G_v2} the phase diagram of the SU($3$) Fermi-Hubbard model on the hypercubic lattice.
We find that two FM phases (FM-I and FM-II) are realized
both below and above one-third filling ($n_{\mathrm{tot}}=1$).
The FM-I phase is characterized by Nagaoka ferromagnetism,
where two of the three flavors are empty and the other is metallic.
By contrast, in the FM-II state,
two of the three flavors are Mott insulating and the other is metallic.
Therefore, both FM-I and FM-II phases are characterized by the flavor-imbalance vector $\bm{v}_3$,
with negative and positive values, respectively.
At the commensurate filling $n_{\mathrm{tot}}=1$, the Mott insulating state is stabilized up to a fairly large temperature, which is proportional to the interaction strength $U$. 
These results indicate that, even in the strong-coupling limit $(U\rightarrow\infty)$,
the Mott insulating state with $m=0$ appears at the commensurate filling,
implying that FM-I and FM-II states cannot be continuously connected.
This should be consistent with the fact that
the particle density of the dominant flavors differ among these states,
as illustrated in Fig.~\ref{fig_G_v2}.

We wish to comment on the effect of the lattice geometry
on the stability of the FM states.
It has been clarified that the FM-I state is stabilized on the hypercubic lattice, whereas it is not stabilized on the Bethe lattice
~\cite{fujii2025itinerant}.
Similar behavior is also observed in the case of $n_{\mathrm{tot}}>1$.
In fact, no peak structure appears in the susceptibility near $n_{\mathrm{tot}} \sim 1.02$ 
(see Appendix~\ref{ap:bethe}) and we could not find the FM state at low temperatures.
This suggests that, even in the FM-II case,
the presence of closed loops in the lattice plays an essential role 
in stabilizing the FM states.

\subsection{Ferromagnetism in the SU($4$) Fermi-Hubbard model}
\begin{figure}
\begin{center}
\includegraphics[width=\linewidth]{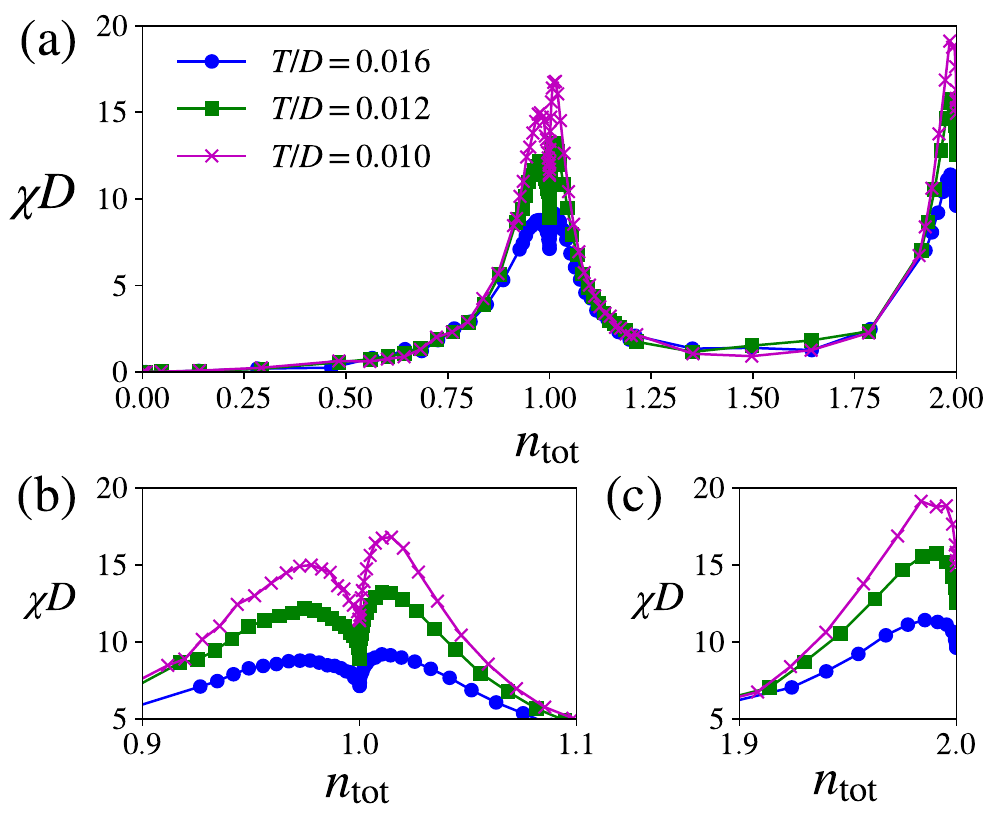}
\end{center}
\caption{
(a) Overall behavior of the susceptibility as a function of $n_{\mathrm{tot}}$
in the $\mathrm{SU}(4)$ Fermi-Hubbard model 
for $U/D = 300$ when $T/D = 0.01, 0.012$, and $0.016$.
(b) [(c)] Enlarged views of the regions around $n_{\mathrm{tot}}=1$ ($n_{\mathrm{tot}}=2$).
}
\label{fig_H_v1}
\end{figure}

We also consider the ferromagnetic instability in the $\mathrm{SU}(4)$ Fermi-Hubbard model
on the hypercubic lattice. 
Figure~\ref{fig_H_v1} shows the filling dependence 
of the magnetic susceptibility 
for $U/D = 300$ when $T/D = 0.01, 0.012$, and $0.016$. 
The SU(4) Fermi-Hubbard model is particle-hole symmetric at half filling,
$n_{\mathrm{tot}} = 2$,
where physical observables are constrained to be symmetric
under particle-hole transformation.
Around quarter filling $(n_{\mathrm{tot}}=1)$, such a symmetry is absent.
As a result, magnetic susceptibilities
are not required to exhibit symmetric behavior with respect to commensurate fillings,
which naturally leads to the observed asymmetry in the peak heights.
We find that magnetic fluctuations develop around $n\sim 0.97$, $1.01$ and $1.99$ 
with decreasing temperature,
indicating that
the FM state is likely to appear even in the SU($4$) model.

\begin{figure}
\begin{center}
\includegraphics[width=\linewidth]{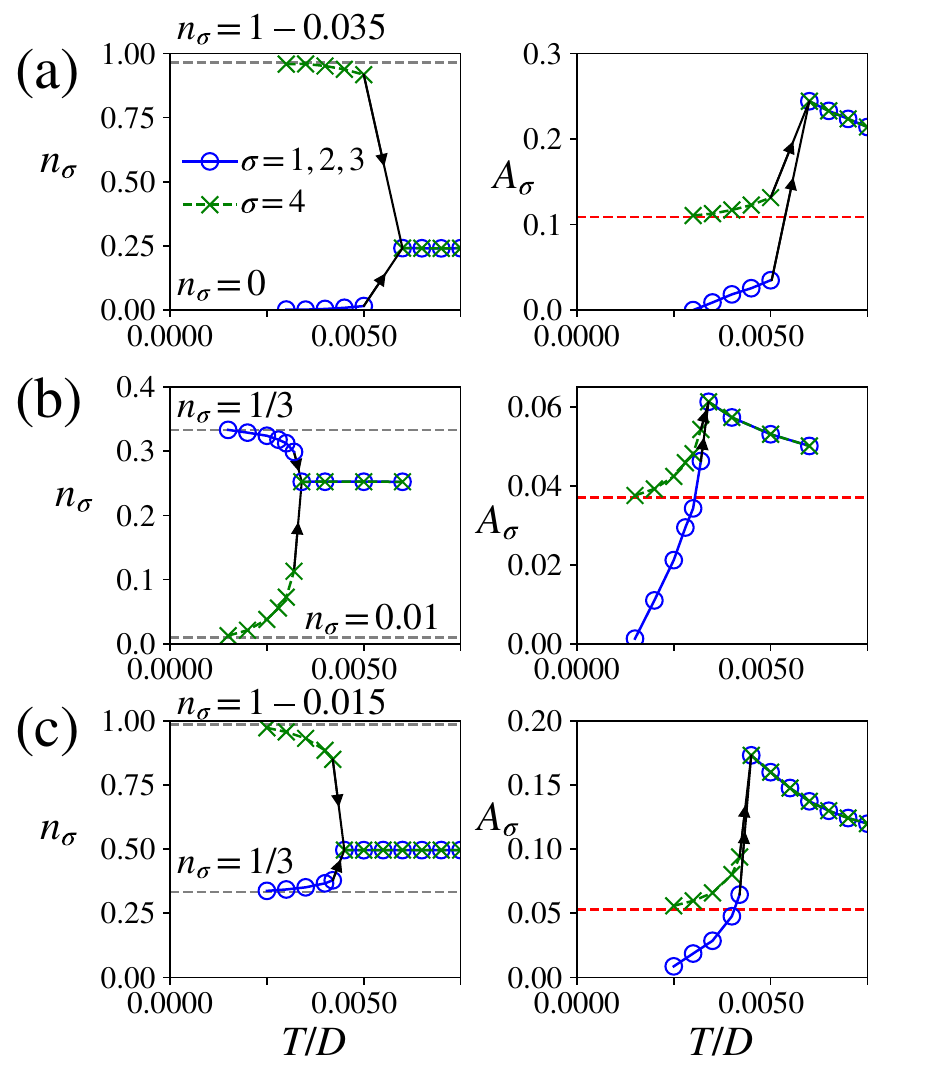}
\end{center}
\caption{
Temperature dependence of the particle density $n_{\sigma}$ (left panels)
and the quantity $A_{\sigma}$ (right panels) in the $\mathrm{SU}(4)$
Fermi--Hubbard model for $U/D = 300$ when (a) $n_{\mathrm{tot}} = 0.965$, (b) $n_{\mathrm{tot}} = 1.01$, and (c) $n_{\mathrm{tot}} = 1.985$, respectively.
Blue circles (green crosses) represent the results for the flavors
$\sigma = 1, 2, 3$ ($\sigma = 4$).
The red dashed lines in right panel of (a), (b), and (c) indicate the value of the noninteracting DOS
corresponding to $n_{4} = 0.965$, $0.01$, and $0.985$, respectively.
Black arrows indicate the evolution of the DMFT solutions upon increasing the temperature $T$.
}
\label{fig_K_v1}
\end{figure}

Figure~\ref{fig_K_v1} shows the temperature dependence of the particle density $n_\sigma$
for (a) $n_{\mathrm{tot}}=0.965$, (b) $n_{\mathrm{tot}} =  1.01$, and (c) $n_{\mathrm{tot}} = 1.985$.
We find a clear particle-density imbalance at low temperatures,
implying that the FM states 
are indeed stabilized at low temperatures.
For the hole-doped region away from quarter filling ($n_{\mathrm{tot}} = 0.965$), the system exhibits a FM phase dominated by a single flavor (FM-I), whereas on the particle-doped side  ($n_{\mathrm{tot}} = 1.01$), a distinct FM state emerges in which three flavors acquire large occupations (FM-II). 
In addition, below half filling ($n_{\mathrm{tot}} = 1.985$), another FM state (FM-III) appears, characterized by the dominant contribution of one flavor while the other three flavors remain partially populated.
We further observe that the four flavors separate into two groups:
three flavors satisfy $n_\sigma=0$ or $1/3$, while the remaining flavor has an intermediate occupation.
This indicates that the FM-I (FM-II and FM-III) state consists of
band (Mott) insulating behavior in the major flavors,
while the minor flavor stays metallic.
This is consistent with the fact that $A_1=A_2=A_3\sim 0$ and $A_4\neq 0$
at low temperatures, as shown in Fig.~\ref{fig_K_v1}.
Therefore, these states with spontaneous symmetry breaking are characterized by 
the flavor-imbalance vector $\bm{v}_4$.
The nature of the first-order magnetic phase transition
is the same as that in the SU(3) case.
Above $T_{c_2}$, the system is in the PM phase,
where physical observables for different flavors become identical,
exhibiting behavior qualitatively similar to that observed in the SU(3) case.
In contrast to $T_{c_2}$, 
determining $T_{c_1}$ where the FM solution suddenly appears
is numerically challenging.
Near $T_{c_1}$, the DMFT self-consistency combined with CTQMC
suffers from strong statistical fluctuations and convergence difficulties,
which become more severe as $N$ increases.
This makes it difficult to reliably stabilize coexisting solutions and
to detect hysteresis behavior. 
Consequently, determining $T_{c_1}$ and obtaining a quantitative finite-temperature phase diagram were
not feasible within our numerical approach.
However, our results demonstrate that,
owing to particle-hole symmetry of the system, six
types of FM states emerge near the commensurate fillings
$n_{\mathrm{tot}} = 1, 2$, and $3$.

\subsection{Ferromagnetism in the SU($N$) Fermi-Hubbard model}\label{sec:d}
\begin{figure}
\begin{center}
\includegraphics[width=\linewidth]{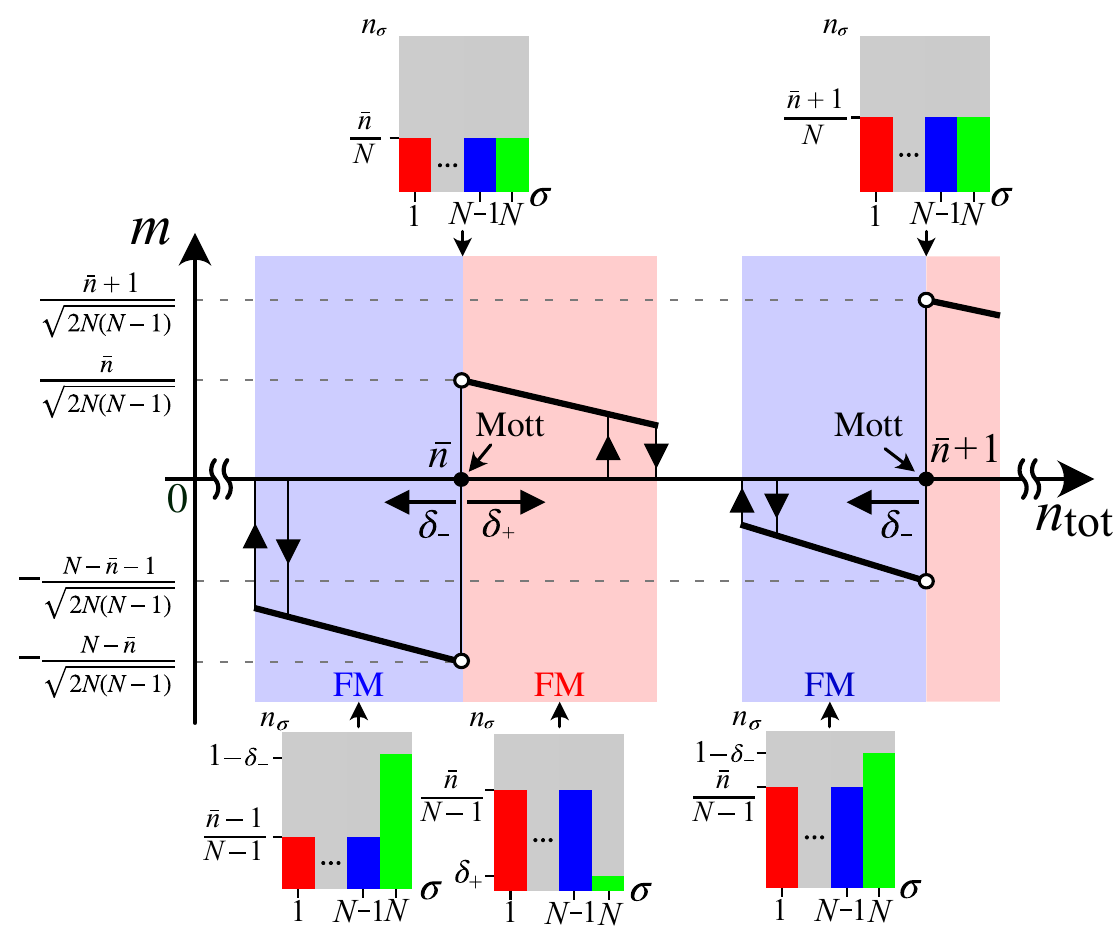}
\end{center}
\caption{
Schematic illustration of the FM states near the commensurate fillings.
The horizontal axis represents the total number of particles $n_{\mathrm{tot}}$, and the vertical axis indicates the magnetization $m$.
Black arrows indicate the evolution of the DMFT solutions
upon varying the total number of particles $n_{\mathrm{tot}}$,
highlighting the coexistence of FM and PM solutions.
}
\label{fig_M}
\end{figure}

We now discuss the nature of the FM ground state
in the SU($N$) Fermi-Hubbard model.
It has been clarified that, near the commensurate fillings $(n_{\mathrm{tot}}=1, 2, \cdots, N-1)$,
magnetic fluctuations are strongly enhanced
and the FM state is stabilized at low temperatures under strong interactions.
Upon hole doping in the system with $1/N$ filling ($n_{\mathrm{tot}}=1$),
the FM state consists of $N-1$ empty flavors and one metallic flavor.
This state is naturally interpreted as a generalization of Nagaoka ferromagnetism.
By contrast, in the other FM states found in our analysis,
$N-1$ flavors are Mott insulating,
while the remaining flavor remains metallic.
Therefore, these ordered states can be characterized by
the spontaneous flavor-selective Mott state with
the flavor-imbalance vector $\bm{v}_N$.
A key point is that $(N-1)$ flavors are band or Mott insulating
while the remaining flavor stays itinerant
due to the lack of interaction-energy cost for hopping.
Since the kinetic energy of this itinerant flavor stabilizes the ordered state,
they may be regarded as
the generalized Nagaoka ferromagnetism extended to the commensurate fillings.

Figure~\ref{fig_M} schematically illustrates the flavor occupations of
the FM states
in the SU($N$) Fermi-Hubbard model with $T = 0$ and $U\rightarrow \infty$.
Let us consider the system around a certain commensurate filling $n_{\mathrm{tot}}=\bar{n}$
with integer $\bar{n}\:(=1, 2, \cdots, N-1)$.
It is well known that, at the commensurate filling $(n_{\mathrm{tot}}=\bar{n})$,
the Mott insulating state is realized, where
the particle density for each flavor is fractional ($n_\sigma=\bar{n}/N$).
When $n_{\mathrm{tot}}=\bar{n}-\delta_-$ with a small positive value $\delta_-$,
the FM state is expected to appear.
In this case,
the particle density of flavors $\sigma = 1, \ldots, N-1$ is $(\bar{n} - 1)/(N-1)$, while
that of flavor $\sigma = N$ is $1 - \delta_-$,
which leads to a negative magnetization.
On the other hand, 
when $n_{\mathrm{tot}}=\bar{n}+\delta_+$ with a small positive value $\delta_+$,
the particle density of flavors $\sigma = 1, \ldots, N-1$ is $\bar{n}/(N-1)$, while
that of flavor $\sigma = N$ is $\delta_+$.
In this case, the magnetization becomes positive.
Since the Mott insulating state is realized at the commensurate filling,
the above two FM states cannot be adiabatically connected to each other.
On the other hand, it may be nontrivial whether
two FM states with the same configuration in the $(N-1)$ flavors, which are located at $n_{\mathrm{tot}} = \bar{n} +\delta_{+}$ and $n_{\mathrm{tot}} = (\bar{n} + 1) - \delta_{-}$,
are adiabatically connected.
Since the occupation of the remaining flavor is either nearly empty or nearly fully occupied,
a sign change in the magnetization naturally occurs.
Furthermore, we have observed that the FM states in the SU($2$)~\cite{kamogawa_ferromagnetic_2019}, SU($3$), and SU($4$) cases
are not stable for large $\delta_-$ and $\delta_+$.
Therefore, we expect that a total of $2(N-1)$ distinct FM states emerge
in the SU($N$) Fermi-Hubbard model.
\section{Conclusions}\label{sec:conclusion}
We have investigated the $\mathrm{SU}(N)$ Fermi-Hubbard model on the hypercubic lattice, 
employing DMFT combined with the CTQMC method. 
In the $\mathrm{SU}(3)$ Fermi-Hubbard model above one-third filling, 
we have found that
the FM state appears at low temperatures under strong interactions,
accompanied by a spontaneous flavor-selective Mott state
where two flavors are Mott insulating and the remaining flavor is metallic.
Extending the analysis to SU($4$), we have identified six
distinct types of FM states away from the commensurate fillings.
We have also discussed the ferromagnetic instability of
the SU($N$) Fermi-Hubbard model.
Overall, our results clarify the origin of ferromagnetism in SU($N$)
Fermi-Hubbard systems and reveal their rich magnetic behavior.
Since $\mathrm{SU}(N)$-symmetric systems 
have been realized in ultracold atomic gases~\cite{desalvo_degenerate_2010,zhang2014spectroscopic,taie2010realization,Fukuhara2007degenerate,hofrichter_direct_2016,pasqualetti2024EOS,pagano2014one,ozawa_antiferromagnetic_2018}, our results can be tested experimentally, e.g., by using quantum gas microscopes~\cite{bakr2009quantum,cheuk2015quantum}.

In our study, we have restricted our analysis to magnetic orders characterized by a spatially uniform order parameter since we have focused on strong particle correlations in the SU($N$) Fermi–Hubbard model away from commensurate fillings as a first step. 
Clarifying how ordered states with enlarged unit cells—such as antiferromagnetic~\cite{jarrell1992hubbard,snoek2008antiferromagnetic,rampon2025magnetic} and canted magnetic orders~\cite{snoek2011canted,sotnikov2013magnetic,Ueda2012Electronic,Gradi2018Correlation}—compete with the FM state remains an important direction for future work and will be addressed elsewhere.

Finally, we briefly comment on the role of dimensionality.
Within the present single-site DMFT framework,
nonlocal spatial correlations are neglected,
and the weakening or destruction of long-range magnetic order due to enhanced spatial
fluctuations in lower dimensions cannot be quantitatively assessed in this work.
On the other hand, for SU($N$) Fermi-Hubbard models,
the existence of Nagaoka ferromagnetism has been rigorously established
in systems with closed loops~\cite{katsura_2013}.
This suggests that, in low-dimensional systems,
ferromagnetic correlations may in principle be observable in experiments with ultracold atomic gases.
However, the corresponding transition temperatures are expected to be low,
and the FM phase is expected to compete with other magnetic phases
such as antiferromagnetically ordered and stripe ordered phases~\cite{partial_Bird,corbozPRL,bohler2025magnetic},
which may make experimental detection more demanding.

\begin{acknowledgments}
This work was supported by JSPS KAKENHI Grants No.\ JP22K03525, JP25H01521, JP25H01398 (A.K.), No.\ JP25K17327 (K.Y.). This work was partly funded by Hirose Foundation, the Precise Measurement Technology Promotion Foundation, the Fujikura Foundation, and the Hiki Foundation, Institute of Science Tokyo. Parts of the numerical calculations were performed in the supercomputing systems in ISSP, the University of Tokyo. Parts of the simulations have been performed using the ALPS libraries~\cite{bauer_alps_2011}.
\end{acknowledgments}

\section*{DATA AVAILABILITY}
The data that support the findings of this article are openly available~\cite{fujii_2026_18610490}.

\appendix
\section{Isotropy of the magnetic susceptibility}\label{ap:iso}
Here, we show that the magnetic susceptibility is isotropic in the PM state. 
For convenience, we present the diagonal generators of $\mathrm{SU}(N)$ in the following form: 
\begin{equation}
  \lambda_{\alpha} = \sqrt{\frac{2}{\alpha(\alpha+1)}} \, I_{\alpha} \oplus (-\alpha) I_1 \oplus \tilde{0}_{N-\alpha-1} 
  \quad (\alpha = 1, \ldots, N - 1),
  \label{eq_Hk}
\end{equation}
where $I_k$ and $\tilde{0}_k$ denote the $k$-dimensional identity 
and zero matrix, respectively. 
This definition satisfies the orthonormality condition 
$\mathrm{Tr}(\lambda_{\alpha} \lambda_{\beta}) = 2\,\delta_{\alpha, \beta}$.
Then, the external field term in an arbitrary direction can be written as
\begin{equation}
  \hat{H}_{\mathrm{ext}} 
  = -\sum_{\alpha=1}^{N-1} h_{\alpha} \hat{S}_{\alpha}.
\end{equation}
Here, the external field used in the main part of the paper 
corresponds to choosing 
$h_{N-1} \neq 0$ and $h_{\alpha} = 0$ for $\alpha = 1, \ldots, N-2$.
The corresponding change in the chemical potential for each flavor induced by the external field 
is given by
\begin{equation}
    (\Delta \bm{\mu})_{\sigma} = -\sum_{\alpha = 1}^{N-1} \frac{h_{\alpha}}{2}(\lambda_{\alpha})_{\sigma,\sigma}.
\end{equation}
When the interaction strength and temperature are fixed, 
the vector $\bm{n} \equiv (n_1,\ldots, n_N)$ depends only on the chemical potential. 
Accordingly, 
it can be expanded as
\begin{equation}
  \bm{n}(\bm{\mu} + \Delta \bm{\mu})
  = 
  \bm{n}(\bm{\mu})
  + 
  \bm{J}_{\bm{n}}(\bm{\mu})\,\Delta \bm{\mu}
  + 
  \mathcal{O}(|\Delta \bm{\mu}|^2),
  \label{eq_Jacobian}
\end{equation}
where 
$[\bm{J}_{\bm{n}}(\bm{\mu})]_{i,j} \equiv
\partial n_i / \partial \mu_j$, 
and the exchange symmetry among flavors ensures that 
the diagonal and off-diagonal components of the Jacobian matrix 
$\bm{J}_{\bm {n}}(\bm{\mu})$ 
take the same value as $d \equiv [\bm{J}_{\bm{n}}(\bm{\mu})]_{i,i}$ and $o \equiv [\bm{J}_{\bm{n}}(\bm{\mu})]_{i\ne j}$, respectively.

The magnitude of the total magnetization can be expressed in terms of the particle densities as
\begin{equation}
  \tilde{m} \equiv \sqrt{\sum_{\alpha=1}^{N-1} m_{\alpha}^2} =
  \sqrt{
    \frac{1}{2}
    \left(
      \sum_{\sigma} n_{\sigma}^2
    \right)
    -
    \frac{1}{2N}
    \left(
      \sum_{\sigma} n_{\sigma}
    \right)^2
  }.
\end{equation}
When the external field is sufficiently weak, using Eq.~\eqref{eq_Jacobian}, we obtain
\begin{equation}
  \tilde{m}^2 
  \simeq
  \frac{(o - d)^2}{2}
  \sum_{\sigma} (\Delta \bm{\mu})_{\sigma}^2
  +
  \frac{1}{2}
  \left\{
    2od + (N - 2)o^2
  \right\}
  \left(
    \sum_{\sigma} (\Delta \bm{\mu})_{\sigma}
  \right)^2.
\end{equation}
From the orthonormality relation of the generalized Gell-Mann matrices, 
the first term satisfies $\sum_{\sigma} (\Delta \bm{\mu})_{\sigma}^2 = h^2 / 2$, where $h \equiv \sqrt{\sum_a |h_a|^2}$ denotes the magnitude of the external field. 
The second term vanishes because the trace of the matrices is zero. 
Therefore, the magnetic susceptibility is given by
\begin{equation}
  \chi = \lim_{h \rightarrow 0} \frac{\tilde{m}}{h} = \frac{|o - d|}{2},
\end{equation}
which is independent of the direction of the external field, 
demonstrating the isotropy of the magnetic susceptibility.
\section{Magnetic susceptibility on the Bethe lattice}\label{ap:bethe}
\begin{figure}
\begin{center}
\includegraphics[width=\linewidth]{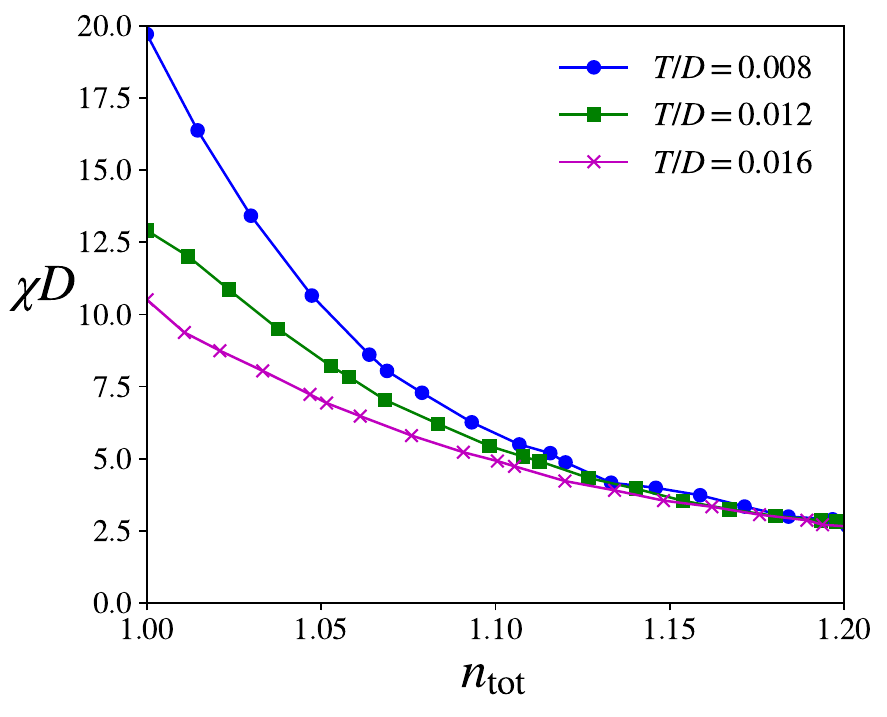}
\end{center}
\caption{
Filling dependence 
of the susceptibility $\chi D$ 
in the $\mathrm{SU}(3)$-symmetric case 
on a Bethe lattice for $U/D = 300$.
Blue, green, and magenta lines 
correspond to $T/D = 0.008$, $0.012$, and $0.016$, respectively.
}
\label{fig_O}
\end{figure}
Here, we investigate magnetic fluctuations in the SU($3$) Fermi-Hubbard model
above one-third filling 
on the Bethe lattice, which has infinite coordination number and provides an exact solution within the DMFT framework. 
In the Bethe lattice, 
the number of nearest-neighbor sites 
is the same for all sites, 
and it does not contain closed loops. 
This implies that the generalized Nagaoka theorem that ensures the existence of the FM state
is not applicable~\cite{nagaoka_1966,katsura_2013}. 
Indeed, it has been suggested that 
below one-third filling, 
the ferromagnetic order 
does not appear 
on the Bethe lattice~\cite{fujii2025itinerant}. 
The noninteracting DOS is given by
\begin{equation}
  \rho(\epsilon) = \frac{2}{\pi D} \sqrt{1 - \left(\frac{\epsilon}{D}\right)^2},
\end{equation}
which leads to the simple form of the Dyson equation for the impurity model as
\begin{equation}
  \mathcal{G}^{\sigma}(i\omega_n)^{-1} 
  = 
  i\omega_n + \mu - \frac{D^2}{4} G_{\mathrm{loc}}^{\sigma}(i\omega_n).
  \label{eq_bethe}
\end{equation}
Combining Eq.~\eqref{eq_bethe} with Eq.~\eqref{weiss_function} yields the self-consistent equation 
in the imaginary-time domain as 
$\Delta^{\sigma}(\tau) = \frac{D^2}{4} G_{\mathrm{loc}}^{\sigma}(\tau)$.

Figure~\ref{fig_O} shows the filling dependence 
of the magnetic susceptibility 
for the $\mathrm{SU}(3)$-symmetric case 
on the Bethe lattice 
above one-third filling. 
In contrast to the hypercubic lattice, 
where the susceptibility shows a peak structure above one-third filling (see Fig.~\ref{fig_A_v1}), 
such nonmonotonic behavior is not observed
on the Bethe lattice for any temperature.
Since the emergence of ferromagnetic order 
is accompanied by an enhancement of magnetic fluctuations, 
this result suggests that 
the FM state does not appear 
above one-third filling on the Bethe lattice, which is similar to the $\mathrm{SU}(2)$ and $\mathrm{SU}(3)$ cases~\cite{kamogawa_ferromagnetic_2019,fujii2025itinerant}. 
These results indicate that 
the emergence of ferromagnetism 
near commensurate fillings 
in the $\mathrm{SU}(N)$ Fermi-Hubbard model strongly depends on the lattice geometry. 

\end{document}